\documentstyle[aaspp4]{article}
 
\lefthead{Elmegreen and Elmegreen}
\righthead{Fractal Structure in Galaxy Star Fields}
\slugcomment{{\it Astronomical Journal}, March 2001, Vol 121}
\begin{document}
 
\title{Fractal Structure in Galactic Star Fields}

\author{Bruce G.~Elmegreen\altaffilmark{1} and Debra Meloy Elmegreen\altaffilmark{2}} 
\altaffiltext{1}{IBM
Research Division, T.J. Watson Research Center, P.O. Box 218, Yorktown
Heights, NY 10598, bge@watson.ibm.com}
\altaffiltext{2}{Department of Physics and Astronomy, Vassar College, 
Poughkeepsie, NY 12604;
e--mail: elmegreen@vassar.edu}

\begin{abstract} The fractal structure of star formation on large scales
in disk galaxies is studied using the size distribution function of
stellar aggregates in kpc-scale star fields. Achival HST images of 10
galaxies are Gaussian smoothed to define the aggregates, and a count of
these aggregates versus smoothing scale gives the fractal dimension.
Fractal and Poisson models confirm the procedure. The fractal dimension
of star formation in all of the galaxies is $\sim2.3$. This is the same
as the fractal dimension of interstellar gas in the Milky Way and nearby
galaxies, suggesting that star formation is a passive tracer of gas
structure defined by self-gravity and turbulence. Dense clusters like
the Pleiades form at the bottom of the hierarchy of structures, where
the protostellar gas is densest. If most stars form in such clusters,
then the fractal arises from the spatial distribution of their
positions, giving dispersed star fields from continuous cluster
disruption. Dense clusters should have an upper mass limit that
increases with pressure, from $\sim10^3$ M$_\odot$ in regions like the
Solar neighborhood to $\sim10^6$ M$_\odot$ in starbursts.
\end{abstract}

keywords: galaxies: star clusters --- ISM: structure ---
stars: formation

\section{Introduction}

Interstellar gas is observed to have a fractal structure ranging from
sub-parsec scales to $>10$ parsec scales in non-self-gravitating clouds
(Crovisier \& Dickey 1983; Green 1993; Vogelaar \& Wakker 1994), from
parsec to $\sim100$ pc scales in self-gravitating clouds (Dickman,
Horvath \& Margulis 1990; Scalo 1990; Falgarone, Phillips, \& Walker
1991; Elmegreen \& Falgarone 1996; Stutzki et al. 1998), and from
$\sim10$ pc scales to $\sim5$ kiloparsec scales in large sections of
galaxies for both the stars (Feitzinger, \& Braunsfurth 1984;
Feitzinger, \& Galinski 1987; Elmegreen \& Efremov 1996; Efremov \&
Elmegreen 1998; Selman \& Melnick 2000;
Elmegreen et al. 2001) and the gas (Stanimirovic et al.
1999; Westpfahl et al. 1999; Keel \& White 2000; Elmegreen, Kim, \&
Staveley-Smith 2001). In many cases, the observed range of scales is
probably a lower limit, because it begins at the scale of resolution
of the instrument and ends at the size of the mapped region. 

The observation of fractal structure in the gas suggests that stars should
form in fractal patterns too if their birth places uniformly follow the
densest regions (see review in Elmegreen et al. 2000). Here we show
evidence for such fractal patterns in the star fields of other galaxies
covering a range of scales from the resolution limit of $\sim10$ parsecs
to giant spiral arm complexes that are several kiloparsecs in length.
Fractal models that are fit to these observations suggest that the
fractal dimension of star formation is around 2.3, which is the same as
for the interstellar gas.

\section{Observations}

We studied optical images of 10 galaxies using Hubble Space Telescope 
archival data. The HST
gives a relatively clear view of star fields in other galaxies because
the factor of 10 improvement in angular resolution over ground-based
images means that distant galaxies can be studied with the same spatial
resolution as conventional images of nearby galaxies but with $\sim100$
times fewer foreground stars.

The galaxies are listed in Table 1. HST archival images of big star
forming regions in these galaxies were convolved with Gaussian point
spread functions in order to blur them to varying degrees. The count of
the number of optical clusters versus the smoothing scale was then
plotted on a log-log plot and the slope was determined. For a fractal
distribution, the slope of such a plot is the fractal dimension, $D$
(Mandelbrot 1983), provided there is no loss of counts from blending.
The slope was determined for the five biggest star fields in NGC 2207,
and for the biggest star fields in 8 other galaxies, with two fields in
NGC 5457. 

\begin{table}
\label{table:gal}
\caption{Galaxies studied}
\begin{tabular}{llcc}
Galaxy&type&Distance (Mpc)&Image scale (pc per WF px)\\
N2207&SAB(rs)bc&35&17\\
2366&IB(s)m	&2.9&1.4\\
3184&SAB(rs)cd&8.7&4.2\\
3423&SA(s)cd	&10.9&5.2\\
4051&SAB(rs)bc&17&8.2\\
4303&SAB(rs)bc&15.2&7.3\\
4449&IBm	&3&1.4\\
5068&SB(s)d	&6.7&3.2\\
5457&SAB(rs)cd	5.4&2.6\\
I2163&SB(rs)c pec&35&17\cr
\end{tabular}
\end{table}

An example of this process is illustrated in Figure 1. This shows six
stages in the smoothing of a 5 kpc-long star field in the south eastern
arm of NGC 2207 (see Elmegreen et al. 2000). We count 75 separate {\it
centers for star formation} (i.e., clusters) in the highest resolution
image, and we count 52, 38, 21, 8, and 2 centers in the five other
images, respectively, which were smoothed in successive steps equal to a
factor of 2 in scale. 

The cluster counts are shown on the left in Figure 2.  The counts for
five star fields
in NGC 2207 are on the top left and the counts for 9 star fields in the
8 other galaxies are on the bottom left. The distribution function for
the number, $n$, of clusters versus scale, $S$, is $n(S)d\log S\propto
S^{-D}d\log S$ for $D= 1.12\pm 0.25$ in the 14 total cases. Thus the fractal
dimension would be $D\sim 1.12\pm 0.25$ without blending. However, the
complexes overlap and blend with each other because of their hierarchical
structure. Thus we have to model this counting process with images of
known fractal dimensions in order to reconstruct the dimension
of the real star fields. 

\section{Models}

Fractal and other models of clusters were made by computer in order to
fit the slope of the observed $n(S)$ relation, and to see 
whether we can tell the difference between a fractal
pattern and completely random pattern, which has a Poisson distribution.
Figure 3 shows sample models before Gaussian smoothing; on the left is
a Poisson distribution of centers, in the middle is a
fractal with $D=1.3$, and on the right is a fractal with $D=2.3.$

The Poisson pattern was made by placing 2048 points on an (x,y) plane with
random positions $x$ and $y$ distributed uniformly between values 0 and 1. 
This is a two-dimensional array, but is equivalent to a three
dimensional array viewed in projection (i.e., random $z$ values
collapsed to the same $z$ value). To
simulate what we already know about clusters, the points were given
finite sizes that have a power-law distribution function comparable to
the observed power law for individual star-cluster sizes, namely
$n(R)d\log R\propto R^{-2.3}d\log R$ (Elmegreen \& Salzer 1999;
Elmegreen et al. 2001). In reality, this intrinsic distribution probably
arises from the same fractal structure that we seek to measure in the
distribution of cluster {\it center positions}, just as the size and
mass distributions of individual molecular clouds display a microcosm of
the same overall fractal structure that is seen on much larger scales in
the distribution of interstellar gas (Elmegreen \& Falgarone 1996).
However, the conventional picture has individual clouds or star-forming
regions with a power law size distribution and a random distribution for
the centers of these regions.  Here we seek to 
disprove this conventional picture
by showing that the center positions are fractal without commenting 
directly on the intrinsic size distribution.

The size distribution used for these models is consistent with the
size-luminosity relation for star clusters, $L\propto R^{2.3}$
(Elmegreen et al. 2001), and with the luminosity distribution function
for clusters and HII regions, $n(L)dL\sim L^{-2}dL$ (Kennicutt, Edgar,
\& Hodge 1989; Battinelli et al. 1994; Elmegreen \& Efremov 1997;
Comeron \& Torra 1996; Feinstein 1997; Oey \& Clarke 1998; McKee \&
Williams 1997). We have commented previously how these relations are
also consistent with a purely fractal distribution, the first giving the
fractal dimension in another way (Pfenniger \& Combes 1994; Larson 1994;
Elmegreen \& Falgarone 1996; Elmegreen
et al. 2001), and the second coming from a hierarchical distribution
with any fractal dimension (Fleck 1996; Elmegreen \& Falgarone 1996).

The model fractal distributions are generated by uniformly selecting
some random number, $N_1$, in the range from 1 to $N$ and then using
this for the number of star-forming regions in the first, or highest,
level in the hierarchy of structures. The (x,y) positions of these $N_1$
regions are then determined uniformly in the interval of position from 0
to 1 using other random variables. Second, we go to the position of each
of these $N_1$ regions and select other random numbers, $N_{2,1}$,
$N_{2,2}, ...$, in the interval from 1 to $N$. These are the number of
level-2 sub-regions associated with each previous region in level 1. For
each level-2 subregion, we find new random positions, but this time
separated from the level-1 positions by a random number in the interval
from 0 to $L<1$, where $L=10^{\log(N)/D}$ for fractal dimension $D$. For
the level-3 positions, we find the number of sublevels first in the same
way, and then choose new positions around each, separated by a random
number in the interval from 0 to $L^2$. With these successively smaller
separations, we make clusters with a fractal dimension,
$D=\log(N)/\log(L)$. This process is continued for 6 levels.

When the selection of fractal positions is finished, we assign each
circle a size randomly distributed according to the function
$n(R)d\log R\propto R^{-2.3}d\log R$, as discussed above. This is done
by solving for $R$ in the equation \begin{equation}
R={{R_{min}}\over{\{1-(1-[R_{min}/R_{max}]^{2.3})\xi\}^{1/2.3}}}
\end{equation} where $\xi$ is a random number uniformly distributed in
the interval from 0 to 1. An image of these circles is then stored on a
$512\times512$ grid. The value of the image is set to 1 inside each
circle, and when two or more circles overlap, the value in the image is
the sum of each contribution. This procedure is consistent with the
approximately constant surface brightness of star complexes that is
implied by the luminosity-size relation given above.

The model images are viewed in Photoshop with different Gaussian
smoothing, stepped by factor-of-2 intervals from the original image. Thus
the smoothing scales are 2, 4, 8, 16, 32, and 64 pixels. The number of
separate regions was counted by eye on each smoothed image.

Figure 2 shows the counts for each image as a function of smoothing
scale. The Poisson maps are steeper than the fractals on these plots
because there is less blending of the small features on the Poisson maps.
This result illustrates the effects of projection mentioned by 
Mandelbrot (1983) and modeled with the shadows of crumpled newspapers by
Beech (1992),
namely, that the dimension of a projected fractal
is approximately one less than the dimension of the full object. 
Figures with low fractal dimensions have the most blending and shallowest
slopes.  The average slopes for the Poisson, $D=2.3$, and $D=1.3$ models
are $-1.72\pm0.04$, $-1.17\pm0.06$, and $-0.75\pm0.09$, respectively.
The slope of the models is about equal to the slope of the observation
for $D=2.3$. 

\section{Discussion}

The distribution of star formation sites in a galaxy is a fractal with
about the same dimension as the fractal interstellar gas. This implies
that stars form from the gas, tracing its structure in a passive way.
This result is not inconsistent with the observation that star formation
occurs in the densest parts of the gas. We add to this observation only the
fact that these densest parts are arranged in space on a fractal
network. Presumably this distribution of star formation sites is the
result of turbulence compression (Elmegreen 1994; Elmegreen 1999;
Rosolowsky et al. 1999; MacLow \& Ossenkopf 2000; Pichardo et al. 2000)
and gravity (Semelin, \& Combes 2000).

The fractal distribution of star formation sites is consistent with the
observation that the total duration of star formation in a region is
always around 2 crossing times, regardless of scale (Elmegreen 2000). It
takes only $\sim1$ crossing time for turbulence to establish the
hierarchy of structures from an initially uniform gas, and it takes
another crossing time on any level for all of the smaller scale
processes, which operate faster, to make their stars. 

Dense star clusters form at the bottom of this hierarchy of gas and
star-formation structures, where the density is large. The maximum
mass of a dense cluster depends on the local pressure and density
as, \begin{equation} M\le 6\times10^3 \left(P/10^8 \;{\rm K\;
cm^{-3}}\right)^{3/2} \left(n/10^5 \;{\rm cm}^{-3}\right)^{-2} .
\end{equation} This comes from the equations $P\sim0.1GM^2/R^4$ and
$n\sim3M/\left(4\pi \mu R^3\right)$ for cloud mass $M$, radius $R$,
core pressure $P$, and mean molecular weight $\mu\sim4\times10^{-24}$
g (Elmegreen 1989; Harris \& Pudritz 1994). A core pressure of $10^8$
K cm$^{-3}$ and an average density of $\sim10^5$ cm$^{-3}$ are chosen
for normalization because these are observed in the Orion regions where
dense clusters form (Lada, Evans, \& Falgarone 1997). The pressure
comes from the density multiplied by the square of the observed velocity
dispersion of $\sim1.5$ km s$^{-1}$. This density of $\sim10^5$ cm$^{-3}$
corresponds to $5.9\times10^3$ M$_\odot$ pc$^{-3}$, and to a final star
density of $\sim10^4$ stars pc$^{-3}$ with 50\% efficiency.  This equation
illustrates why the galactic or ``open'' clusters in our Milky Way disk,
which are born with stellar densities like this, tend to be smaller than
several thousand solar masses.  Higher ambient interstellar pressures
should lead to higher cloud core pressures and the formation of more
massive clusters with the same and higher densities.

Most star formation seems to occur in dense clusters, although many of
these may disperse soon after birth (Kroupa 2000). Even so, the
distribution of young stars should still be fractal in an overall
fractal gas because the velocity dispersion of each cluster is small
compared to the turbulent velocity dispersion of the larger region
around it. This means that the timescale for the larger scale in the
hierarchy of structures is always shorter than the time for a dense
cluster to expand to this large scale. Because of this, cluster
evaporation and dispersal on the small scale should not smear out the
fractal pattern that is continuously established by turbulence and
self-gravity on the large scale.

\newpage
\begin{figure}
\caption{(see jpg file) Six levels of Gaussian smoothing of a star-forming patch
in the galaxy NGC 2207. The number of pixels in the Gaussian
smoothing function is shown in each panel.
The number of objects is plotted as a function
of this smoothing length in Fig. 2. }
\end{figure}

\begin{figure}
\vspace{3.5in}
\includegraphics{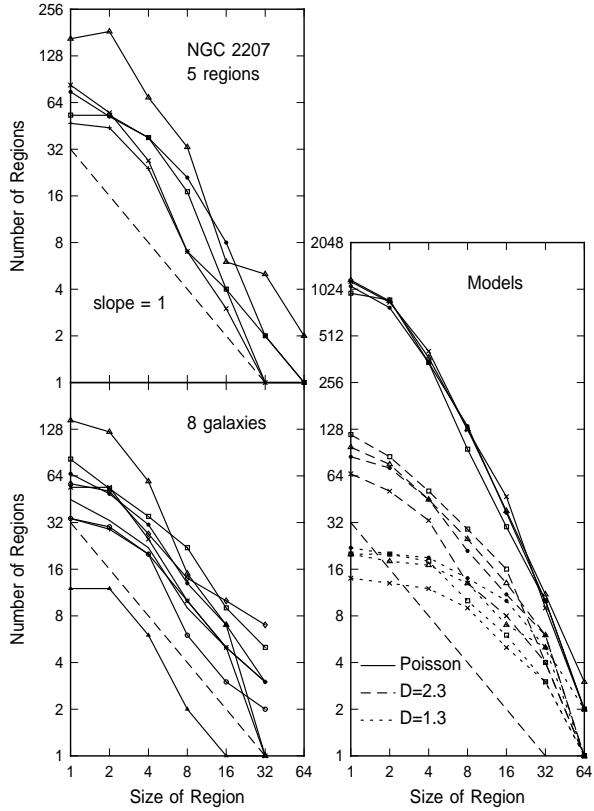}
\caption{The number of star-forming concentrations is plotted as a function
of the smoothing scale for five regions in NGC 2207 on the top left,
for 9 regions in 8 galaxies on the lower left, and for three types
of models on the right. The dashed line has a slope of $-1$ on this
log-log plot. The observations are best fit by a fractal with a dimension
of 2.3, as shown by the dashed line on the right. }
\end{figure}

\begin{figure}
\caption{(see jpg file) Three models for the spatial distribution of star-forming
regions. The sizes of the regions have the observed power law
distribution.  These models are Gaussian smoothed to varying
degrees to make the counts shown on the right in Fig. 2.}
\end{figure}

\end{document}